# Transmission properties of periodic nonlinear structures containing left handed materials.

Ravi S. Hegde, Herbert G. Winful

*Department of Electrical Engineering and Computer Science, University of Michigan, Ann Arbor, MI 48109*

**Abstract.** We study the transmission properties of a nonlinear periodic structure containing alternating slabs of a nonlinear right handed material and a linear left handed material. We find that the transmission associated with the zero averaged- refractive- index gap exhibits a bistable characteristic that is relatively insensitive to incident angle. This is in contrast to the nonlinear behavior of the usual Bragg gap.

Negative index media, first proposed by Veselago [1], have attracted a great deal of theoretical and experimental interest in recent years. The phenomenon of negative refraction was first experimentally demonstrated at microwave frequencies using artificial metamaterials [2] and there are currently attempts to extend this behavior to the optical regime [3-5]. A periodic assembly of alternating positive index and negative index materials [6-8] has been found to possess exotic properties. It has been shown that in the mixed type periodic structure, a new kind of gap results [9], when the average refractive index of the structure becomes zero. The zero-n gap has very unusual properties in that it is robust to scaling and disorder [9]. The gap exhibits an omnidirectional feature [10], which makes it of great interest for applications requiring a wide field of view. These properties of the zero-n gap are quite different from those of the usual Bragg gap. The nonlinear response of positive index Bragg structures was first studied by Winful, et al, who showed that intensity-tuning of the transmission bandgap can lead to hysteresis and optical bistability [11]. In this paper we investigate the nonlinear transmission of a positive-negative periodic structure. We find that the zero-n gap exhibits a hysteretic response which is relatively insensitive to input angle, in contrast to the behavior of the Bragg gap.

We consider a structure in which the positive index medium exhibits Kerr-type nonlinearity. The negative index medium is taken to be linear although nonlinearity in its electric and magnetic susceptibilities can be easily included. We ignore absorption in the structure, which can also be readily included within this formalism The periodic structure (as in Fig. 1) consists of N unit cells occupying the region z = 0, D, translationally invariant in the x-y plane and bounded on both sides by free space. Each cell is formed of two films film 1 and film 2 with thicknesses d1 and d2. Film 1 is a positive index material, while film 2 is a negative index or left handed material. The dielectric permittivity described by eqn (1) is taken to exhibit field intensity dependence. We ignore such effects for the magnetic permeability as done in [12]. The implicit assumption in this formulation is that the negative index layer permits the use of an effective index for the frequency range of interest, which may not be a valid approximation in all cases.

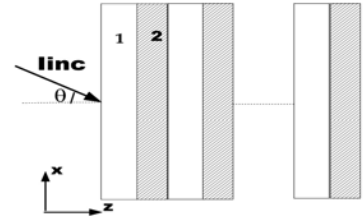

**FIGURE 1**. Schematic of the nonlinear periodic structure.

$$\hat{\epsilon}_i(\omega, z) = \epsilon'_i(\omega, z) + \epsilon_{iNL}(|\mathbf{E}|^2)$$
$$(i = 1, 2) \quad (1)$$

We solve for TE polarized fields of the form,

$$E(x, z) = E(z) e^{ik_o(\beta x - ct)}$$
$$k_o = \omega/c$$
$$\beta = \sin(\theta) \quad (2)$$

where θ is the angle which the incident wave vector makes with the x axis. Now setting $\zeta = k_o z$. The electric field is in general governed by eqns. (3-4).

$$\frac{d^2 E}{d\zeta^2} + p^2 E + \mu_i \epsilon_{iNL}(|\mathbf{E}|^2) = 0 \quad (3)$$

$$p^2 = \mu_i(\epsilon'_i(\omega, z) - \beta^2) \quad (4)$$

The solutions for the most general case have to be obtained by numerical integration of eqn (3). In case of a Kerr-type nonlinearity, the $\epsilon_{iNL}(|E|^2)$ term in eqn.

(1) can be replaced by $\gamma|E|^2$. To obtain the transmission of the structure we specify the E field and its derivative at the output face, successively integrate eqn (3) all the way to the input end. At the interfaces of two films we apply the following boundary conditions.

$$E_i = E_j$$
$$\mu_j \frac{dE}{d\zeta}\bigg|_i = \mu_i \frac{dE}{d\zeta}\bigg|_j \quad (5)$$

In case we have a linear film, we can instead use the following transmission matrix to propagate the fields. Here the subscript refers to the film number in the unit cell. If the film is a negative index type, then a negative value of p is used.

$$\begin{bmatrix} \cos(p_i k d_i) & (-1/p)\sin(p_i k d_i) \\ p\sin(p_i k d_i) & \cos(p_i k d_i) \end{bmatrix} \quad (6)$$

The procedure outlined above has been implemented using MATLAB. The numerical integration being handled by a standard Runge-Kutta solver. Our structure has d1 = d2 =10mm. Film 1 is a positive index material with ε' = 2, μ = 1. The positive layer exhibits a Kerr-type nonlinearity modeled by γI. Film 2 is a negative index material whose relative permittivity and permeability variations (f is measured in GHz) are of the form

$$\epsilon'_2(f) = 1.6 + \frac{40}{0.81 - f^2}, \quad (7)$$

$$\mu_2(f) = 1 + \frac{25}{0.814 - f^2} \quad (8)$$

The dispersion relations and the linear transmission spectrum (for N = 16,32) are shown in Fig. 2

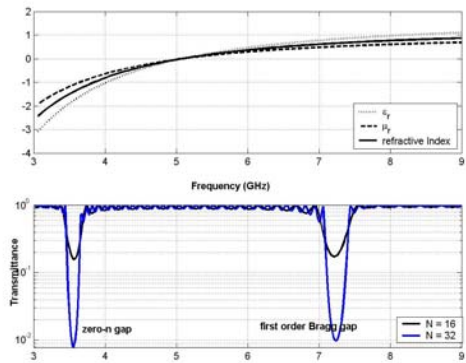

**FIGURE 2**. Material parameters of the negative layer and the transmittance of the structure as a function of frequency at normal incidence, N = 16, 32.

To study the shift in transmission spectra, we employ a root solver to maintain constant input intensity. For a defocusing γI$_{inc}$ = -0.03, Fig 3 shows gap tuning for the zero-n gap (θ = 0°, 45°). Two things to be noted are the dependence on incident angle and the asymmetry in the shifts. The zero-n gap results at a frequency when the following condition is satisfied.

$$\sqrt{\epsilon_1(f)}\sqrt{\mu_1(f)}d_1 = \sqrt{\epsilon_2(f)}\sqrt{\mu_2(f)}d_2$$

The structure is thus expected to show omnidirectionality. However in a finite structure, the gap center does vary with the incident angle and the number of layers. This can be attributed to the fact that in a finite structure the end reflections influence the overall transmission. The only asymmetry in the problem is due to the dispersion relations in eqns (7-8) and this is what leads to the observed asymmetry in the transmission shifts.

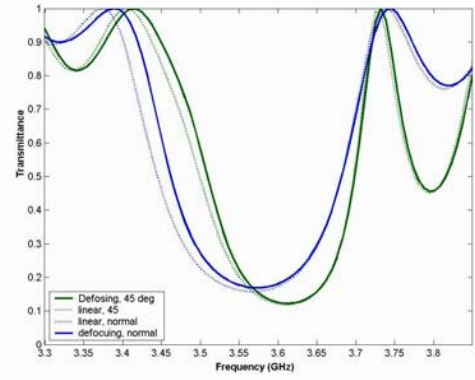

**FIGURE 3** Gap tuning for the zero-n gap at θ = 0°, 45°. γI$_{inc}$ = -0.03, N = 16.

Next we examine the hysteresis effects that the transmission shows as a function of input intensity. We vary γI$_{inc}$ and observe the transmission at a particular frequency and incident angle. The hysteresis curves for detuning near the first order Bragg gap (Fig. 4) are very sensitive to incidence angle. In contrast for detuning near the zero gap center (f near 3.55 GHz), the hysteresis curves (Figs. 5,6) for a wider range of incidence angles are seen to follow each other closely. This is the result of the omnidirectionality feature. Due to the asymmetry in the shifts for detuning to the right and left of the zero-n gap, a relatively larger nonlinearity is essential to observe hysteresis in the former case. The phenomenon of negative refraction is always accompanied by dispersion [1] and hence it will play a key role in nonlinear gap tuning. By changing dispersion in the negative index material, the gap tuning and hysteresis behavior can be controlled.

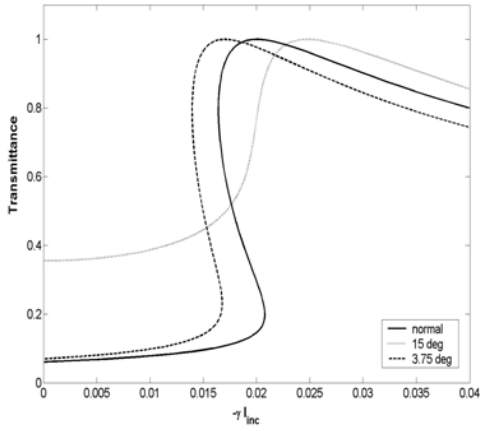

**FIGURE 4.** Hysteresis behavior of transmittance as a function of a defocusing $\gamma I_{inc}$ for detuning near the first order Bragg gap (f = 7.15 GHz, N = 32.) for incident angles $\theta = 0°$, 3.75°. Note that for $\theta = 15°$, hysteresis is not observed.

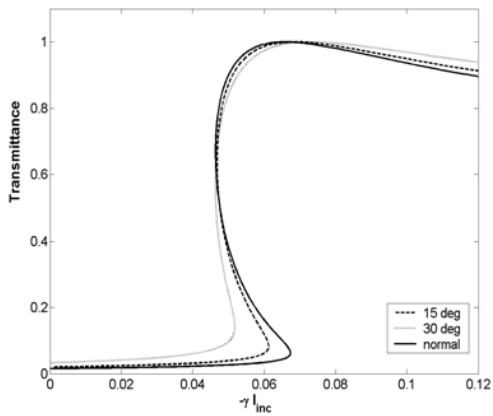

**FIGURE 5.** Hysteresis behavior of transmittance as a function of a defocusing $\gamma I_{inc}$ for detuning to the left of the zero-n gap (f = 3.51 GHz, N = 32.) for incident angles $\theta = 0°$, 15° and 30°

In summary, we introduce a new kind of periodic nonlinear structure, containing a left handed material. A general method to analyze nonlinear transmission effects for this class of structures is presented. As in the case of a linear left handed bandgap structure, a zero-n gap is observed. Gap tuning and hysteresis are observed in this structure. Differences between the nonlinear transmission properties between the zero-n gap and the first order Bragg gap are demonstrated. In the scheme presented, it will be an interesting extension to consider periodic nonlinear structures with defect layers.

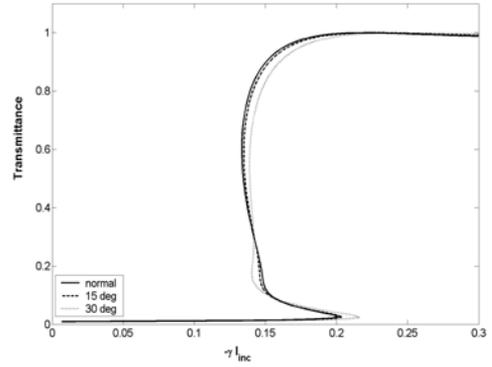

**FIGURE 6.** Hysteresis behavior of transmittance as a function of a defocusing $\gamma I_{inc}$ for detuning to the right of the zero-n gap (f = 3.58 GHz, N = 32.) for incident angles $\theta = 0°$, 15° and 30°